\documentclass[onecolumn]{IEEEtran}

\ifCLASSINFOpdf
   \usepackage[pdftex]{graphicx}
   \graphicspath{{../pdf/}{../jpeg/}}
   \DeclareGraphicsExtensions{.pdf,.jpeg,.png}
\else
   \usepackage[dvips]{graphicx}
   \graphicspath{{../eps/}}
   \DeclareGraphicsExtensions{.eps}
\fi
\usepackage{diagbox}
\usepackage{color}
\usepackage{amsmath}
\usepackage{bm}
\usepackage{amssymb}
\usepackage{multirow}   
\usepackage{makecell}
\usepackage{rotating}
\usepackage{setspace}
\doublespace
\begin{document}
%
\title{A Lightweight Deep Network for Efficient CSI Feedback in Massive MIMO Systems}
%
%
%

\author{Yuyao~Sun,~
        Wei~Xu,~
        Le~Liang,~
        Ning~Wang,~
        Geoffery~Ye~Li,~\IEEEmembership{Fellow,~IEEE,}
        and~Xiaohu~You,~\IEEEmembership{Fellow,~IEEE}
\thanks{Y. Sun, W. Xu, and L. Liang are with the National Mobile Communications Research Laboratory, Southeast University, Nanjing 210096, China (e-mail: yy.sun@seu.edu.cn; wxu@seu.edu.cn; lliang@gatech.edu).}
\thanks{N. Wang is with the Department of Electrical Engineering, Zhengzhou University, Zhengzhou, Henan 450001, China (e-mail: ienwang@zzu.edu.cn).}
\thanks{G. Y. Li is with the Department of Electrical and Electronic Engineering, Imperial College London, London, UK (e-mail: geoffrey.li@imperial.ac.uk).}
\thanks{X. You is with the National Mobile Communications Research Laboratory, Southeast University, Nanjing 210096, China, and also with Purple Mountain Laboratories, Nanjing 211111, China (e-mail:  xhyu@seu.edu.cn).}
}

\maketitle

\begin{abstract}
To fully exploit the advantages of massive multiple-input multiple-output (m-MIMO), accurate channel state information (CSI) is required at the transmitter. However, excessive CSI feedback for large antenna arrays is inefficient and thus undesirable in practical applications. By exploiting the inherent correlation characteristics of complex-valued channel responses in the angular-delay domain, we propose a novel neural network (NN) architecture, namely ENet, for CSI compression and feedback in m-MIMO. Even if the ENet processes the real and imaginary parts of the CSI values separately, its special structure enables the network trained for the real part only to be reused for the imaginary part. The proposed ENet shows enhanced performance with the network size reduced by nearly an order of magnitude compared to the existing NN-based solutions. Experimental results verify the effectiveness of the proposed ENet.
\end{abstract}


%
\IEEEpeerreviewmaketitle

\section{Introduction}
%
%
%
%
\IEEEPARstart{M}{assive} multiple-input multiple-out (m-MIMO) is a key enabling technology of the 5G wireless communication systems due to its high spectral efficiency. Nevertheless, a necessary yet challenging prerequisite for m-MIMO is to obtain high-dimensional channel state information (CSI) at the transmitter. In frequency division duplex (FDD) systems, downlink CSI is acquired at the user equipment (UE) and then reported to the base station (BS) through a feedback channel. The large number of antennas in m-MIMO systems results in a high-dimensional channel matrix, which imposes a heavy burden on the feedback link.

The application of deep learning in wireless communication systems has been a research hot spot in recent years and successful deployment of neural networks (NNs) in CSI feedback \cite{ref1}, end-to-end communication \cite{ref2}, and channel estimation \cite{ref3} has been realized. Benefiting from considerable progresses of NN in computer vision and wireless communication \cite{ref4}, a convolutional NN-based approach, namely CsiNet, has been proposed in \cite{ref1}, exhibiting superior performance in CSI compression by treating the angular-delay domain channel matrix as a sparse 2D image. Particularizing the design of CsiNet in various scenarios, e.g., with high mobility and temporal correlation, a series of works then developed methods using NN for the m-MIMO CSI feedback. Existing works are mainly classified into two categories. One category uses more sophisticated network architectures, including the joint convolutional residual JC-ResNet \cite{ref5}, CsiNet+ \cite{ref6} and multi-resolution CRNet \cite{ref7}. The other category is to exploit extra correlation information. Recurrent NN architectures were employed in \cite{ref8} and \cite{ref9} to exploit the temporal correlation of CSI in time-varing channels. In \cite{ref10}, the uplink channel was used as an auxiliary input of the NN to assist the reconstruction of compressed downlink CSI, which assumes reciprocity between the uplink and downlink channels. In \cite{ref11}, a cooperative recovery network, named CoCsiNet, was proposed to cut back on the feedback overhead by exploiting the shared information of UEs in proximity. A new module, named Anci-block, was devised in \cite{ref12} by exploiting the visualized characteristics of the angular-delay domain channel matrix.

Because the channel gains are complex-valued, existing methods typically stacked the real and imaginary parts as an entire real-valued input to fit the requirement of popular NN design. However, it is natural to conjecture that there exists some kind of similarity between the two parts even though the real part is proven statistically independent of the imaginary part if the channel is symmetrically complex Gaussian distributed. In this article, we observe that the real and imaginary parts of the complex-valued channel matrix in fact share almost the same correlation even though there exist distinct correlations in the angular and delay domains. Based on this finding, we devise an efficient NN, namely ENet, for m-MIMO CSI feedback with a substantially reduced size. We adopt individual compression strategies for the angular and delay domains to exploit domain-specific correlations. Instead of stacking the real and imaginary parts together, as in existing NN architectures for CSI feedback, we propose the scheme of one-part training and two-parts deployment to take advantage of the similarity in the correlations of the real and imaginary parts of CSI. Particularly in our proposed ENet, only the real part of the CSI matrix is utilized for the network training while both the real and imaginary parts are compressed and fed back using the same trained network. While the network size is reduced by nearly an order of magnitude, the ENet exhibits brilliant performance compared to the existing NN-based methods.

\section{System Model}

We consider the downlink of an FDD m-MIMO system, in which ${N_{\rm{t}}}$ antennas are deployed at the BS and a single antenna is installed at the UE. Orthogonal frequency division multiplexing (OFDM) with ${N_{\rm{c}}}$ subcarriers is used, where the received signal at the $n$th subcarrier is expressed as

\begin{equation}
\label{1}
y_n  = {\bf{h}}_n^H {\bf{v}}_n x_n  + z_n,
\end{equation}
where ${{\bf{h}}_n} \in {\mathbb{C}^{{N_t} \times 1}}$ denotes the channel vector of the $n$th subcarrier, ${{\bf{v}}_n} \in {\mathbb{C}^{{N_t} \times 1}}$ is the precoding vector, ${x_n} \in \mathbb{C}$ is the transmit symbol, and ${z_n} \in \mathbb{C}$ is the additive noise.

The downlink channel matrix in the spatial-frequency domain is denoted by ${\bf{H}} = {[{{\bf{h}}_1} \cdots {{\bf{h}}_{{N_{\rm{c}}}}}]^H} \in {\mathbb{C}^{{N_{\rm{c}}} \times {N_{\rm{t}}}}}$, which requires a total feedback of ${N_{\rm{c}}}{N_{\rm{t}}}$ complex-valued scalars if no compression is adopted. This can be extremely large in m-MIMO systems where a typical antenna array size is about a few hundreds. In order to compress the CSI for feedback through a bandwith-limited feedback channel, we transform ${\bf{H}}$ to the angular-delay domain, which explicitly presents the channel sparsity \cite{ref13}. By using a 2D discrete Fourier transform (DFT), the CSI in the angular-delay domain is expressed as

\begin{equation}
\label{2}
{{\bf{H}}_{\rm{a}}} = {{\bf{F}}_{\rm{c}}}{\bf{H}}{\bf{F}}_{\rm{t}}^H,
\end{equation}
where ${{\bf{F}}_{\rm{c}}} \in {\mathbb{C}^{{N_{\rm{c}}} \times {N_{\rm{c}}}}}$ and ${{\bf{F}}_{\rm{t}}} \in {\mathbb{C}^{{N_{\rm{t}}} \times {N_{\rm{t}}}}}$ are DFT matrices. As ${{\bf{H}}_{\rm{a}}}$ contains only values in a small delay duration \cite{ref1}, we focus on the first ${N_{\rm{cc}}}$ rows of ${{\bf{H}}_{\rm{a}}}$, denoted by ${{\bf{H}}_{\rm{s}}} \in {\mathbb{C}^{{N_{\rm{cc}}} \times {N_{\rm{t}}}}}$, for CSI compression and feedback.

Further compression is still necessary because the amount of feedback, ${N_{\rm{cc}}}{N_{\rm{t}}}$, can still be very large with a large ${N_{\rm{t}}}$. Fortunately, there are more to be exploited to design an efficient m-MIMO CSI feedback architecture if we go deeper into the channel structure in the angular-delay domain. On the one hand, the correlation in the angular domain behaves differently from that in the delay domain. On the other hand, the real part and imaginary part of the CSI matrix are discovered to share similar correlations.

Based on the above two findings, we design an efficient NN architecture for the m-MIMO CSI feedback. The proposed ENet consists of an Encoder and a Decoder. The Encoder is responsible for generating the compressed representation of ${{\bf{H}}_{\rm{s}}}$ in terms of ${{\bf{H}}_{\rm{R}}}$ and ${{\bf{H}}_{\rm{I}}}$, where ${{\bf{H}}_{\rm{R}}}$ and ${{\bf{H}}_{\rm{I}}}$ are the real and imaginary parts of ${{\bf{H}}_{\rm{s}}}$. The Encoder is denoted by

\begin{equation}
\label{3}
{{\bf{s}}_{\rm{R}}} = {f_{\rm{EN}}}({\bf{H}}_{\rm{R}})\;\; {\rm{and}} \;\; {{\bf{s}}_{\rm{I}}} = {f_{\rm{EN}}}({\bf{H}}_{\rm{I}}),
\end{equation}
where ${{\bf{s}}_{\rm{R}}}$ and ${{\bf{s}}_{\rm{I}}}$ are $M$-dimensional codewords of the real and imaginary parts of the compressed CSI, and $f_{\rm{EN}}( \cdot )$ is the compression operation of the Encoder. Defining $N \buildrel \Delta \over = {N_{\rm{cc}}}{N_{\rm{t}}}$, the compression ratio is $\gamma  = M/N$. At the other side, the Decoder is used to reconstruct ${\bf{H}}_{\rm{R}}$ and ${\bf{H}}_{\rm{I}}$ from the compressed codeword. It yields

\begin{equation}
\label{4}
\widehat{\bf{H}}_{\rm{R}} = {f_{\rm{DE}}}({{\bf{s}}_{\rm{R}}})\;\; {\rm{and}} \;\;\widehat{\bf{H}}_{\rm{I}} = {f_{\rm{DE}}}({{\bf{s}}_{\rm{I}}}),
\end{equation}
where $f_{\rm{DE}}( \cdot )$ is the reconstruction process by the Decoder.

\section{Channel Characteristics and ENet}
In this section, we elaborate the new architecture of the proposed ENet for the m-MIMO CSI feedback, which mainly exploits the two findings of the inherent nature of the m-MIMO channel matrix: 1) the difference of angular-delay domain correlations, 2) the similarity in the correlations of the real and imaginary parts of CSI. Before introducing the details of ENet, we first characterize the correlation of ${\bf{H}}_{\rm{s}}$ in the following.
\begin{figure}[t]
\centering
\includegraphics[width=0.8\textwidth]{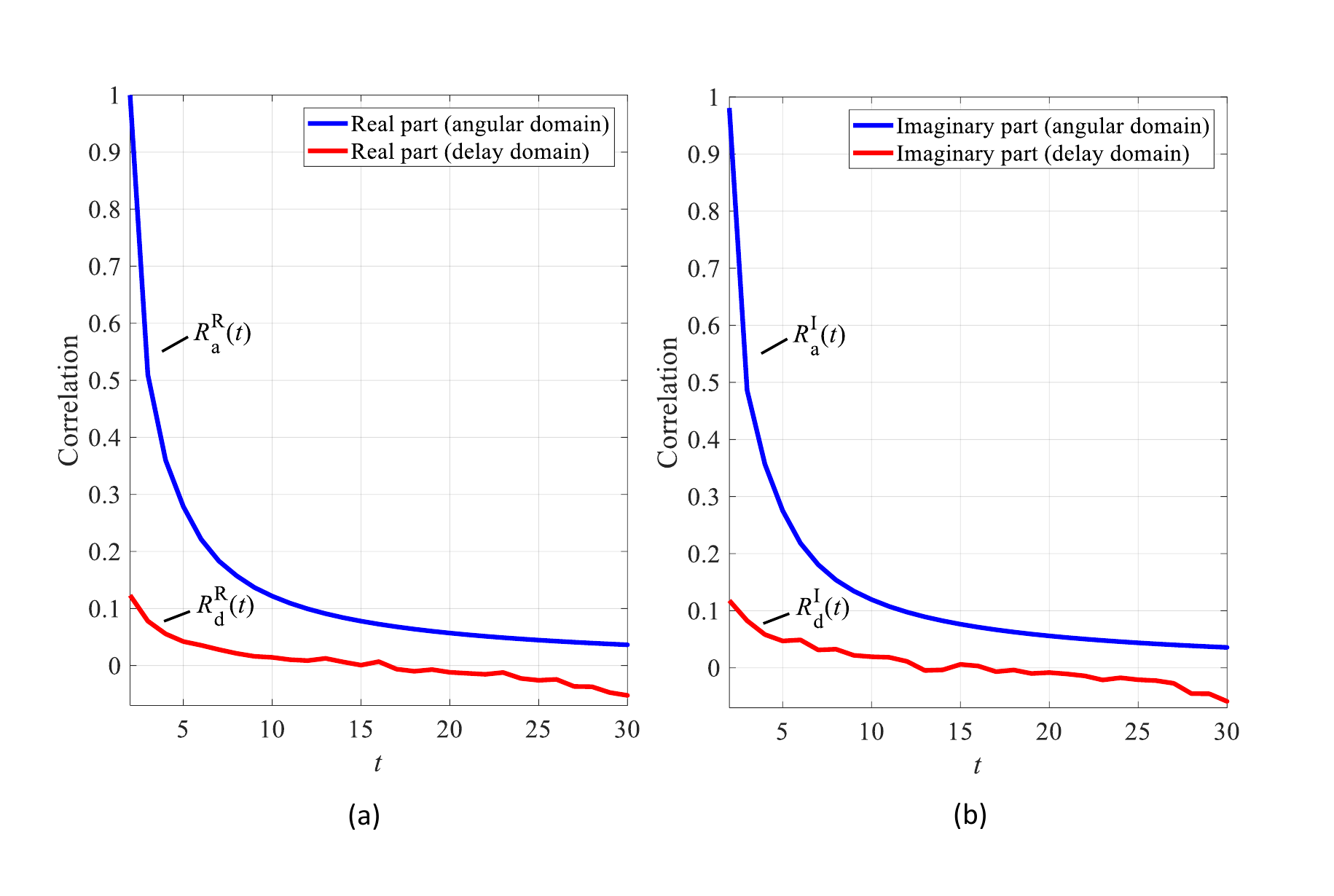}
\caption{(a) Correlation of ${\bf{H}}_{\rm{R}}$. (b) Correlation of ${\bf{H}}_{\rm{I}}$.}
\end{figure}

\subsection{Correlation Difference and Similarity}
The channel correlations between the two ends of the communication link usually depend on the scatterers in the propagation paths. In the delay domain, a resolvable path typically consists of multiple unresolvable paths coming from the same scatterer. Thanks to the large number of antennas in m-MIMO systems, high-resolution in the angular domain is available, which enables unresolvable paths in the delay domain to be resolved in the angular domain. Therefore, a strong correlation can be observed in the adjacent angles. However, the resolvable paths in the delay domain most probably come from different scatterers and a weak correlation is therefore observed in the delay domain. The correlation of ${{\bf{H}}_{\rm{R}}}$ in the angular domain is defined as

\begin{equation}
\label{5}
R_{\rm{a}}^{\rm{R}}(t_{\rm{a}}) = \frac{1}{{N_{\rm{cc}}}}{\mathbb{E}}\left[ {\sum\limits_{i = 1}^{N_{\rm{cc}}} {\left( {{{\bf{H}}_{\rm{R}}} \left( {i,1} \right){{\bf{H}}_{\rm{R}}} \left( {i,1 + t_{\rm{a}}} \right)} \right)} } \right],
\end{equation}
where $t_{\rm{a}} = 1, \cdot  \cdot  \cdot ,{N_{\rm{t}}}  - 1$ represents the interval in the angular domain. The correlation of ${{\bf{H}}_{\rm{I}}}$ in the angular domain, $R_{\rm{a}}^{\rm{I}}(t_{\rm{a}})$, is similarly defined. The correlation of ${\bf{H}}_{\rm{R}}$ in the delay domain is defined as

\begin{equation}
\label{6}
R_{\rm{d}}^{\rm{R}} (t_{\rm{d}}) = \frac{1}{N_{\rm{t}}}{\mathbb{E}}\left[ {\sum\limits_{i = 1}^{N_{\rm{t}}} {\left( {{{\bf{H}}_{\rm{R}}} \left( {1,i} \right){{\bf{H}}_{\rm{R}}} \left( {1 + t_{\rm{d}},i} \right)} \right)} } \right],
\end{equation}
where $t_{\rm{d}} = 1, \cdot  \cdot  \cdot ,{N_{\rm{cc}}}  - 1$ represents the interval in the delay domain. The correlation of ${\bf{H}}_{\rm{I}}$ in the delay domain, $R_{\rm{d}}^{\rm{I}}(t_{\rm{d}})$, is defined in a similar way. Without loss of generality, we set $t = t_{\rm{a}} = t_{\rm{d}}$ in the subsequent discussion, where $1 \le t < \min \left( {{N_{{\rm{cc}}}},\, {N_{\rm{t}}}} \right)$.

\begin{figure*}[t]
\centering
\includegraphics[width=1\textwidth]{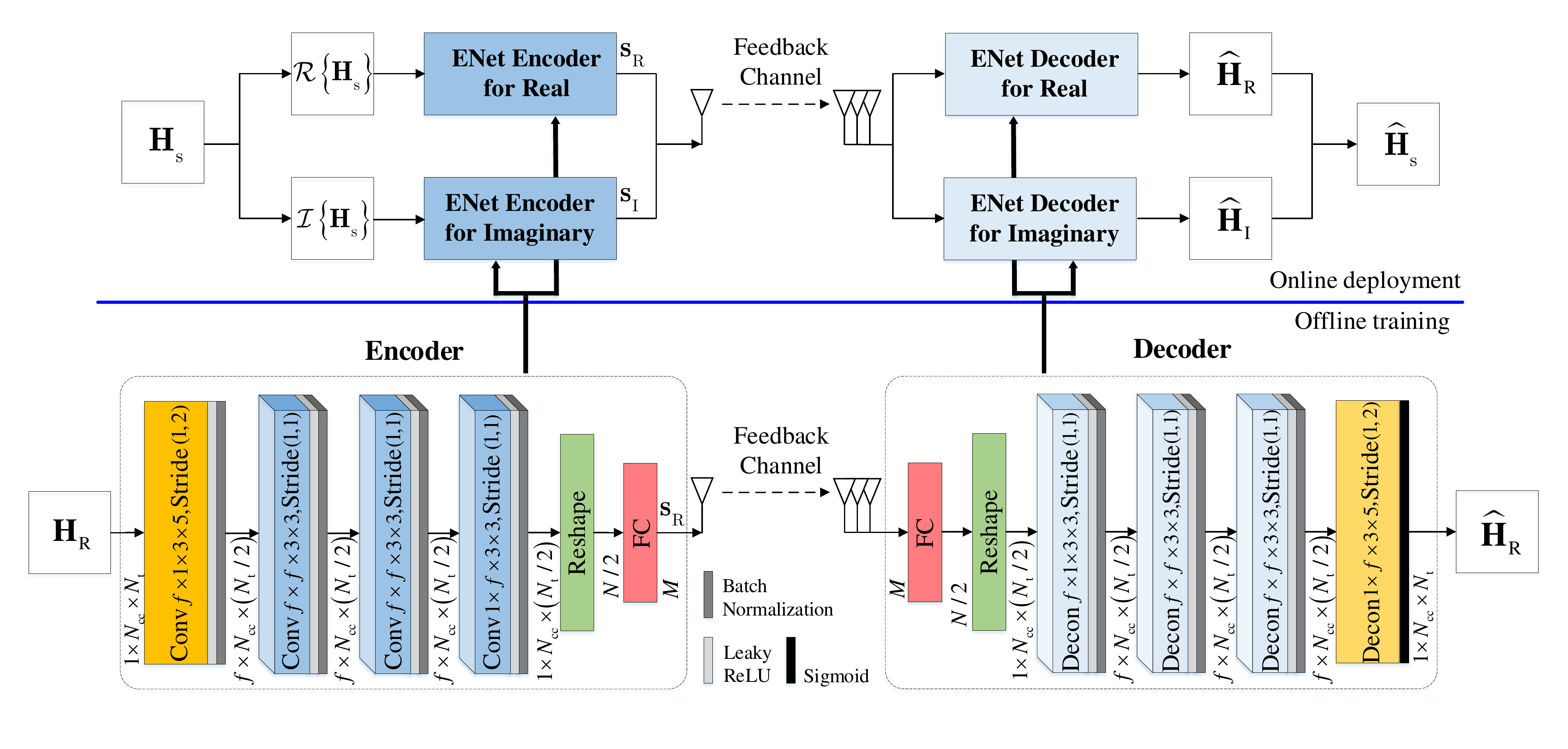}
\caption{Proposed architecture of ENet.}
\end{figure*}

On the other hand, there exists another feature of the statistics of the CSI that is elaborated in the following theorem, proved in Appendix A.

\emph{Theorem 1:} For a channel with the probability density function satisfying (13) and (15), the correlation of the real part of the CSI values is equal to the correlation of the imaginary part, that is $R_{\rm{a}}^{\rm{R}}(t)=R_{\rm{a}}^{\rm{I}}(t)$ and $R_{\rm{d}}^{\rm{R}}(t)=R_{\rm{d}}^{\rm{I}}(t)$, where $1 \le t < \min \left( {{N_{{\rm{cc}}}},\, {N_{\rm{t}}}} \right)$.

To illustrate the correlation difference and similarity, we depict in Fig. 1 the correlation of 150,000 CSI samples of ${{\bf{H}}_{\rm{R}}}$ and ${{\bf{H}}_{\rm{I}}}$ generated from the COST 2100 indoor channel model \cite{ref14}. From the figure, there is a strong correlation in the angular domain whereas a weak correlation exists in the delay domain. In addition, correlation similarity exists between the real and imaginary parts of the complex-valued CSI matrix in the angular domain and similar phenomenon can be observed in the delay domain.

\subsection{The Proposed ENet}

Autoencoder-based NNs for m-MIMO CSI feedback \cite{ref1}, \cite{ref5}-\cite{ref12} utilize convolutional layers in the encoder to extract features before a fully-connected layer, where the disadvantage lies in that the compression is done by the fully-connected layer alone. Note that a single fully-connected layer results in a sharp dimension reduction, leading to excessive impairment of the CSI information. Moreover, the parameters of the fully-connected layer constitutes the majority of the entire network at low compression ratios, which remains large if no compression is done before the fully-connected layer. Taking the correlation difference into consideration, this disadvantage can be well addressed. From the perspective of information theory, the domain with a stronger correlation, i.e., the angular domain, can be compressed to a greater extent. Therefore, it is natural to adopt different compression strategies for the angular and delay domains. Our ENet compresses the CSI matrix in the angular domain before the fully-connected layer.

Stacking the real and imaginary parts as an entire real-valued input is the method that most existing deep learning (DL)-based approaches adopt for complex-valued CSI feedback \cite{ref1}, \cite{ref5}-\cite{ref9}, \cite{ref11}, \cite{ref12}. However, the correlation similarity established in Theorem 1 makes it feasible that only the real part of the CSI matrix is trained while the trained network can be directly applied to the imaginary part. With the input and output size reduced to half of the stacked CSI, the total number of network parameters can be saved by at least a half, resulting in a lightweight network with better performance that is easier to train.

By exploiting the similarity in correlations of the real and imaginary part of CSI and the correlation difference in the angular and delay domains, we propose the ENet for the m-MIMO CSI compression and feedback. Fig. 2 displays the architecture of ENet in detail. We use the three-dimensional values, e.g., $1 \times {N_{\rm{cc}}} \times {N_{\rm{t}}}$, to represent the depth, width, and height of the input tensor, respectively. The four-dimensional values, e.g., $f \times 1 \times 3 \times 5$, denote the number, depth, width, and height of the convolution and deconvolution kernels, respectively. The two-dimensional values, $\left(1,2 \right)$, represent the convolution and deconvolution stride for the width and height of the input tensor, respectively. Since a stronger correlation exists in the angular domain, we use convolution with the stride 2 to compress the CSI in this dimension. It is revealed that the convolution kernel with the size 5 in such a stride produces good performance. The reason of a larger stride convolution employed only in the first layer of the Encoder is that the correlation distribution of ${\bf{H}}_{\rm{R}}$ and ${\bf{H}}_{\rm{I}}$ is deterministic, whereas the correlation distribution may change after the first layer. Correspondingly, we use a deconvolution with stride 2 in the last layer of Decoder to recover CSI in the angular domain. For the other layers and in the delay domain, we use convolutions with stride 1 and convolution kernels with size 3 to process the data stream.

The ENet uses a symmetric architecture for the CSI compression and recovery. After the first layer in Encoder, the size of the CSI matrix in the angular domain is compressed from $N_{\rm{t}}$ to ${N_{\rm{t}}}/2$. Then, two identical convolutional layers are placed in sequence to strengthen network expression capability. Experiments validate that two additional layers enhance the network performance efficiency. A convolution of size $f\times 3 \times 3$ is then used to compress the feature size to $1 \times {N_{\rm{cc}}} \times \left({N_{\rm{t}}}/2\right)$. At the end of the Encoder, the feature maps is reshaped into a vector before a fully-connected layer which is utilized to generate the compressed codeword, ${{\bf{s}}_{\rm{R(I)}}}$.

The Decoder function in ENet serves as an inverse operation of Encoder, which reconstructs the original real and imaginary parts of the CSI matrix, i.e., ${\bf{H}}_{\rm{R}}$ and ${\bf{H}}_{\rm{I}}$, from the received codeword. Except for the last deconvolutional layer in the Decoder, we use batch normalization and the Leaky Rectified Linear Unit (Leaky ReLU) activation function for all convolutional and deconvolutional layers. The batch normalization is used to speed up the convergence of network training and the Leaky ReLU function is chosen as

\begin{equation}
\label{7}
{\rm{LeakyReLU}}\left( x \right) = \left\{ \begin{array}{l}
x, \qquad x \ge 0,\\
0.3x,\ \ x < 0.
\end{array} \right.
\end{equation}
For the last layer in the Decoder, we use a sigmoid activation function to constrain the output values into [0, 1].

The correlation difference and similarity not only makes ENet a reasonable data compression and recovery method, but also grants a lightweight architecture for the m-MIMO CSI feedback. With the help of the correlation similarity, the parameters of the fully-connected layer decrease from $2N \times 2\gamma N + 2\gamma N$ to $N \times \gamma N + \gamma N$, which is a significant reduction in the number of parameters. Further with the correlation difference, the number of parameters drops to $\left( N/2 \right) \times \gamma N + \gamma N$.

Furthermore, we introduce a tunable parameter, $f$, the number of kernels, to control the complexity of ENet. For performance-oriented applications, a large $f$, e.g., $f=32$, is recommended, while for complexity-limited applications, a small $f$, e.g., $f=16$, is preferred.

In order to train ENet, we adopt the end-to-end method and use the mean-squared error (MSE) loss function as

\begin{equation}
\label{8}
L=\frac{1}{T}\sum\limits_{i = 1}^{T}{\left\| {f_{\rm{DE}} (f_{\rm{EN}} ({\bf{H}}_{\rm{R}}[i]  ) ) - {\bf{H}}_{\rm{R}}[i] } \right\|_2^2 },
\end{equation}
where $T$ is the total number of training samples and $\left\|  \cdot  \right\|_2$ is the Euclidean norm. The network training objective is to minimize the MSE of the reconstructed $\widehat{\bf{H}}_{\rm{R}}[i]$ and the corresponding ground-truth values of ${\bf{H}}_{\rm{R}}[i]$.

\section{Experimental Results}
This section presents the performance evaluation of the proposed ENet. A total of 150,000 CSI samples generated with the COST 2100 indoor channel model \cite{ref14} at 5.3 GHz is used to validate our method and we divide the training, validation and test sets to contain 100,000, 30,000, 20,000 samples, respectively. In practical applications, the training data can be obtained either through computer simulations by generating CSI samples according to 3GPP channel models or by conducting offline channel measurements. $N_{\rm{t}}=32$ antennas is placed at the BS with a uniform linear array (ULA) and $N_{\rm{c}}=1024$ subcarriers are assumed. After the angular-delay domain transformation, $N_{\rm{cc}}=32$ rows of ${\bf{H}}_{\rm{s}}$ in the delay domain are reserved.  ${\bf{H}}_{\rm{R}}$ and ${\bf{H}}_{\rm{I}}$ are normalized. The Adam optimizer is used for the network training with a batch size of 1,000.

\begin{table}[t]
\caption{Total Number of Parameters}
\begin{center}
\begin{tabular}{p{0.18\textwidth}<{\centering}|p{0.08\textwidth}<{\centering}|p{0.08\textwidth}<{\centering}|p{0.08\textwidth}<{\centering}|p{0.08\textwidth}<{\centering}}
\hline
$\gamma $ & 1/4 & 1/16 & 1/32 & 1/64 \\
\hline
CsiNet [1] & 2.10M & 0.53M & 0.27M & 0.14M\\
CRNet-cosine [7] & 2.11M & 0.53M & 0.27M & 0.14M\\
ENet $\left( f=16 \right)$& 0.27M & 0.08M & 0.04M  & 0.03M\\
ENet $\left( f=32 \right)$& 0.30M & 0.11M & 0.07M  & 0.06M\\
\hline

\end{tabular}
\end{center}
\end{table}

\begin{figure}[t]
\centering
\includegraphics[width=0.65\textwidth]{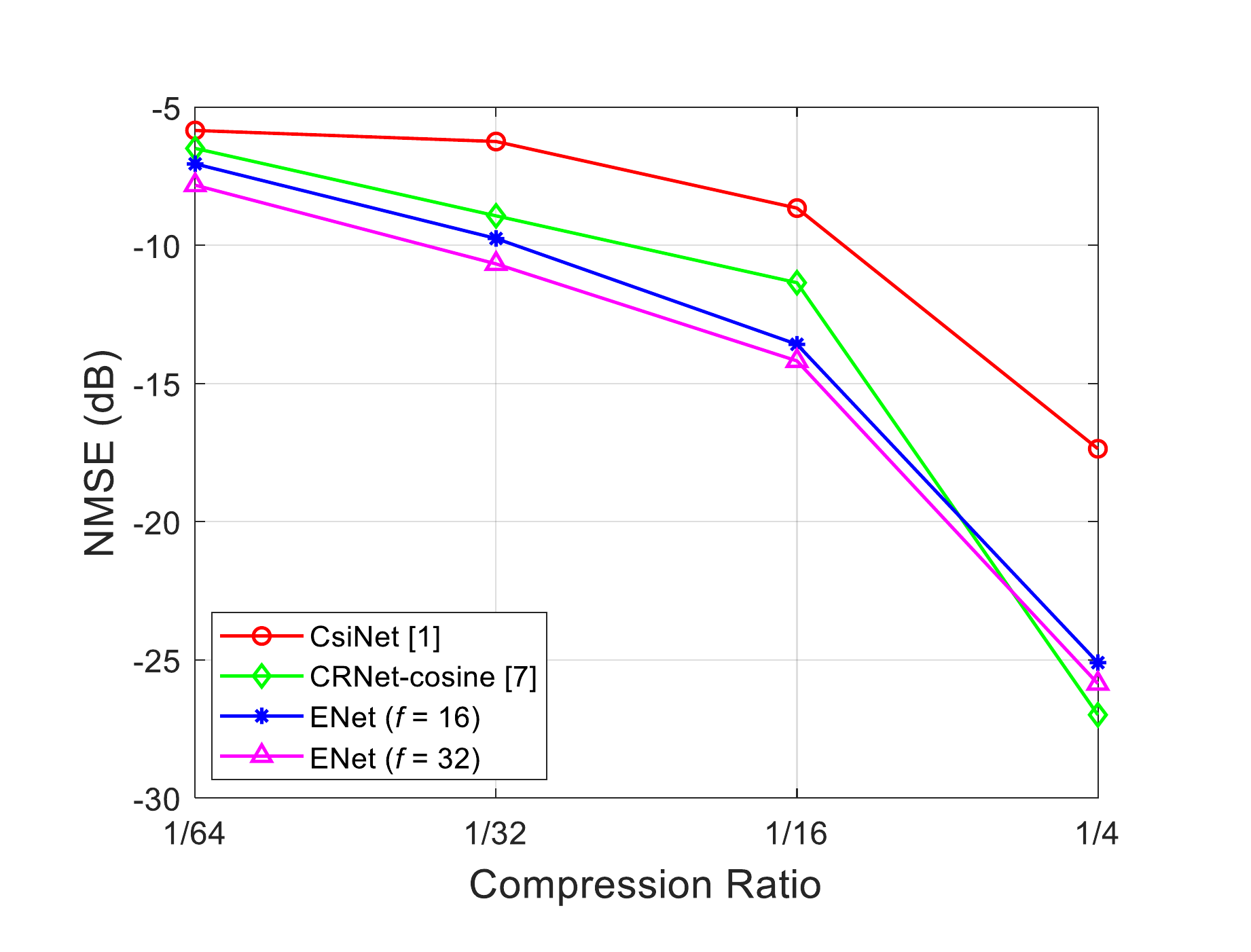}
\caption{NMSE performance at different compression ratios.}
\end{figure}

\begin{figure}[t]
\centering
\includegraphics[width=0.8\textwidth]{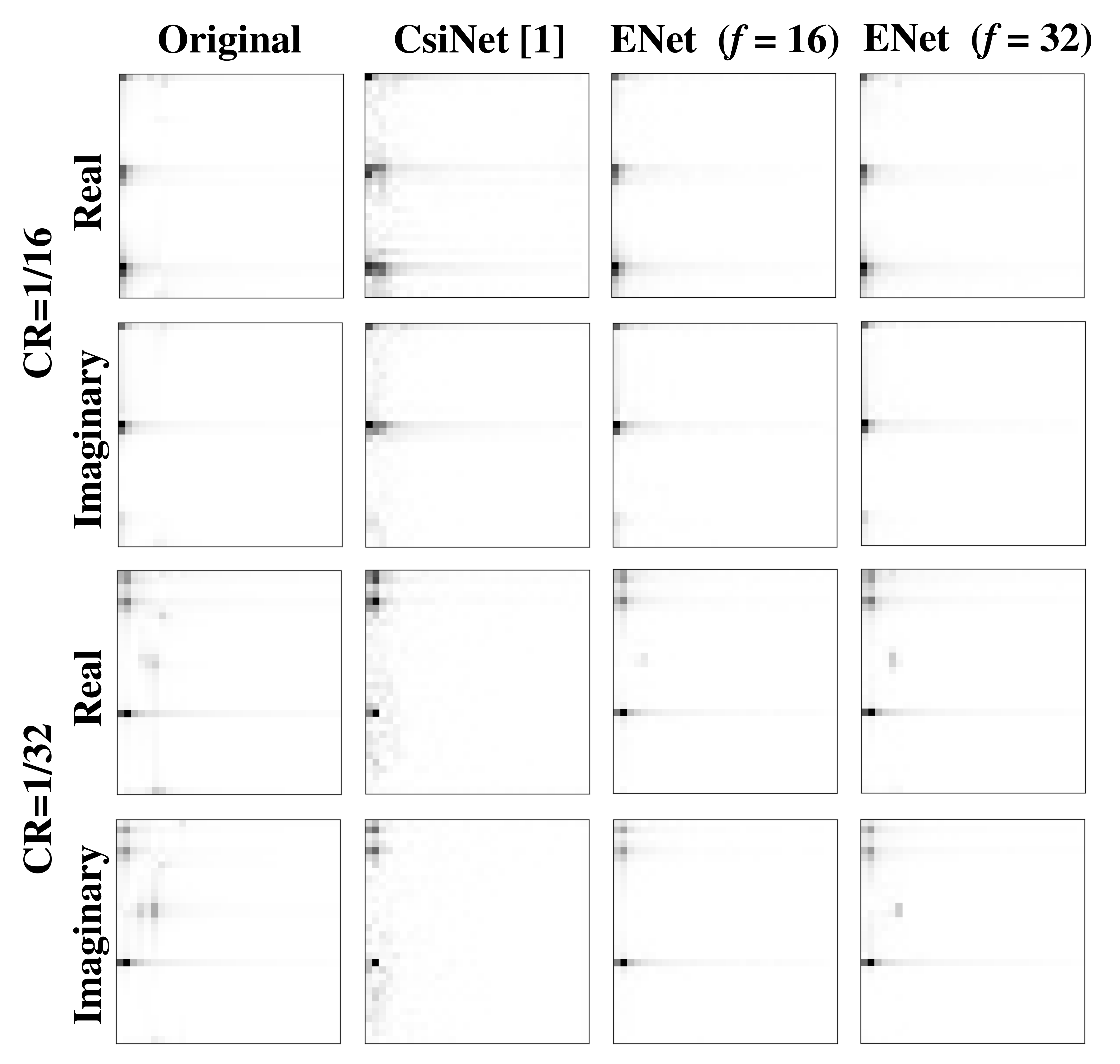}
\caption{Reconstruction visualization at different compression ratios.}
\end{figure}

We compare the network complexity of CsiNet \cite{ref1}, CRNet-cosine \cite{ref7} and our proposed ENet with respect to the number of network parameters in Table I. The parameters of the convolutional layer, the deconvolutional layer, and the fully-connected layer account for the majority of the complexity of the NN, and thus we focus on these components.
From Table I, ENet is a parameter-saving NN compared to both CsiNet and CRNet-cosine. For $f=16$, ENet reduces the number of network parameters by over 80\% at all compression ratios. For $f=32$, 60\% parameters are saved. Fewer parameters not only mean memory-saving, but also require fewer data samples for training and can alleviate overfitting.

We use the normalized MSE (NMSE) defined as,

\begin{equation}
{\rm{NMSE}} = \mathbb{E}\left\{ {\left\| {\widehat{\bf{H}}_{\rm{R(I)}}  - {\bf{H}}_{\rm{R(I)}} } \right\|_2^2 }/{{\left\| {{\bf{H}}_{\rm{R(I)}} } \right\|_2^2 }} \right\},
\end{equation}
to evaluate the CSI reconstruction performance. Fig. 3 compares the NMSE performance by CsiNet \cite{ref1}, CRNet-cosine \cite{ref7}, and ENet, where the performance of CsiNet and CRNet-cosine is tested under ${\bf{H}}_{\rm{s}}$, and the performance of ENet is the average NMSE tested for both ${\bf{H}}_{\rm{R}}$ and ${\bf{H}}_{\rm{I}}$. From the figure, ENet outperforms CsiNet at all compression ratios and CRNet-cosine at high compression ratios. This is due to the distinct compression strategies designed upon the correlation difference. For better illustration, we visualize the reconstructed real and imaginary part in Fig. 4, where the strength of a pixel represents the magnitude of the channel values. For both $f=16$ and $32$, ENet is able to recover the CSI in a more accurate manner than CsiNet at all compression ratios while $f=32$ preserves more subtle features than $f=16$. We also summarize the NMSE performance tested under ${\bf{H}}_{\rm{R}}$ and ${\bf{H}}_{\rm{I}}$ in Table II. From the table, ENet achieves similar performance for both ${\bf{H}}_{\rm{R}}$ and ${\bf{H}}_{\rm{I}}$, which validates the effectiveness of our finding in Theorem 1 on the correlation similarity.

\begin{table}[t]
\caption{NMSE Comparison under Different Test Sets}
\begin{center}
\begin{tabular}{p{0.06\textwidth}<{\centering}|p{0.06\textwidth}<{\centering}|p{0.12\textwidth}<{\centering}|p{0.12\textwidth}<{\centering}}
\hline
\multirow{2}{*}{$\gamma$} &
\multirow{2}{*}{$f$} &
\multicolumn{2}{c}{Test Set NMSE (dB)} \\
\cline{3-4}   
& &${\bf{H}}_{\rm{R}}$ &${\bf{H}}_{\rm{I}}$ \\
\hline
\hline
\multirow{2}{*}{$\frac{1}{4}$} & 16 &-25.09 &-25.11\\
&32 &-25.82 &-25.90\\

\hline
\multirow{2}{*}{$\frac{1}{16}$} & 16 &-13.58 &-13.57\\
&32 &-14.18 &-14.17\\

\hline
\multirow{2}{*}{$\frac{1}{32}$} & 16 &-9.75 &-9.74\\
&32 &-10.66 &-10.67\\

\hline
\multirow{2}{*}{$\frac{1}{64}$} & 16 &-7.03 &-7.06\\
&32 &-7.80 &-7.83\\

\hline
\end{tabular}
\end{center}
\end{table}

\section{Conclusion}

In this article, we have proposed a DL approach, named ENet, for m-MIMO CSI compression and feedback based on the angular-delay domain channel characteristics. We have shown that by utilizing the correlation difference in the angular and the delay domains and the similarity between correlations in the real and imaginary parts of CSI under some mild conditions, we can significantly reduce the size of the NN while still achieving desirable performance. Moreover, such correlation similarity across real and imaginary domains as demonstrated in \emph{Theorem 1} can be useful for other NN designs for complex-valued calculations.

\begin{appendices}
\section{proof of \emph{Theorem 1}}

For an angular-domain channel response ${a_{{\theta}}}{e^{j{\theta}}}$ with independent magnitude ${a_{{\theta}}}$ and phase $\theta$, the correlation of the real part is calculated as

\begin{equation}
\begin{aligned}
{R_{\rm{re}}} &= {\mathbb{E}}\left[ {{a_{{\theta _1}}}  \cos {\theta _1} \cdot {a_{{\theta _2}}} \cos {\theta _2}} \right] \\
&={\mathbb{E}}\left[ {{a_{{\theta _1}}}  {a_{{\theta _2}}}} \right] \cdot \mathbb{E}\left[ {\cos {\theta _1}  \cos {\theta _2}} \right],
\end{aligned}
\end{equation}
where the equality is due to the independence of the magnitude and the phase. Similarly, the correlation of the imaginary part equals

\begin{equation}
{R_{{\rm{im}}}} = \mathbb{E}\left[ {{a_{{\theta _1}}}  {a_{{\theta _2}}}} \right] \cdot \mathbb{E}\left[ {\sin {\theta _1}  \sin {\theta _2}} \right].
\end{equation}
Let us consider the second term of (10). It follows

\begin{equation}
\begin{aligned}
&\mathbb{E}\left[ {\cos {\theta _1}  \cos {\theta _2}} \right] = \iint \cos {\theta _1}  \cos {\theta _2}  f\left( {{\theta _1},{\theta _2}} \right){\mathrm{d}}{\theta _1}{\mathrm{d}}{\theta _2}\\
&= \int{\cos {\theta _1}  f\left( {{\theta _1}} \right)  \left[ {\int {\cos {\theta _2}  f\left( {{\theta _2}|{\theta _1}} \right){\mathrm{d}}{\theta _2}} } \right]{\mathrm{d}}{\theta _1}},
\end{aligned}
\end{equation}
where $f\left( \cdot \right)$ is the probability density function. Assuming that channel phase $\theta$ follows

\begin{equation}
f\left( {{\theta}} \right) = f\left( {{\theta} + \frac{\pi }{2} } \right).
\end{equation}
Defining ${\theta _1}^\prime  \buildrel \Delta \over = {\theta _1} + \frac{\pi }{2}$, we rewrite (12) as

\begin{equation}
\begin{aligned}
&\!\!\mathbb{E}\left[ {\cos {\theta _1}  \cos {\theta _2}} \right] \\
\!\!=&\!\! \int \!\!{\cos\! \!\left(\! {{\theta _1}^\prime \!\!\! -\! \frac{\pi }{2} }\! \right) \!\! {f_{{\theta _1}}}\!\!\!\left(\! {{\theta _1}^\prime \!\!\! - \!\frac{\pi }{2} }\! \right)
\!\!\left[\! {\int\!\! {\cos {\theta _2}  f\!\!\left(\! {{\theta _2}|{\theta _1}^\prime \!\!\! - \!\frac{\pi }{2} }\! \right)\!\!{\rm{d}}{\theta _2}} }\! \right]\!\!{\rm{d}}{\theta _1}^\prime } \\
\!\!=&\!\! \int\!\! {\sin  {{\theta _1}^\prime }   {f_{{\theta _1}}}\!\!\left( {{\theta _1}^\prime } \right)\!\!  \left[ {\int\!\! {\cos {\theta _2} f\!\!\left( {{\theta _2}|{\theta _1}^\prime  - \frac{\pi }{2} } \right)\!\!{\rm{d}}{\theta _2}} } \right]\!\!{\rm{d}}{\theta _1}^\prime }.
\end{aligned}
\end{equation}
Assuming that channel phase $\theta$ further satisfies

\begin{equation}
\label{ddf}
{f_{{\Theta _2}|{\Theta _1}}}\left( {\theta _2}|{\theta _1} \right)={f_{{\Theta _2}|{\Theta _1}}}\left( {\theta _2 + \alpha}|{\theta _1+\alpha} \right),
\end{equation}
where $\alpha \in \mathbb{R}$. In particular, for a channel phase $\theta$ that follows uniform distribution, which holds in an ideal and popular channel model, (13) and (15) hold. Let ${\theta _2}^\prime  \buildrel \Delta \over = {\theta _2} + \frac{\pi }{2}$. We further have

\begin{equation}
\begin{aligned}
&\!\! \mathbb{E}\left[ {\cos {\theta _1}  \cos {\theta _2}} \right] \\
&\!\!\!\!\!\!\!\! =\!\!\! \int\!\! {\sin \!{\theta _1}^\prime  \!{f_{{\theta _1}}}\!\!\!\left({{\theta _1}^\prime } \right)\!\! \left[\! {\int \!\! {\cos\!\! \left(\! {{\theta _2}^\prime\!\!\!  -\! \frac{\pi }{2} }\! \right)\!\! f\!\!\left( \!{{\theta _2}^\prime \!\!\! -\! \frac{\pi }{2} |{\theta _1}^\prime \!\!\! -\! \frac{\pi }{2} }\! \right)\!{\rm{d}}{\theta _2}^\prime } }\! \right]\!\!{\rm{d}}{\theta _1}^\prime } \\
& \!\!\!\!\!\!\!\!=\!\!\! \int\!\! {\sin {\theta _1}^\prime   {f_{{\theta _1}}}\!\!\left( {{\theta _1}^\prime } \right) \!\! \left[ {\int\!\! {\sin {\theta _2}^\prime   {f_{{\theta _2}|{\theta _1}}}\left( {{\theta _2}^\prime |{\theta _1}^\prime } \right){\rm{d}}{\theta _2}^\prime } } \right]{\rm{d}}{\theta _1}^\prime } \\
& \!\!\!\!\!\!\!\!= \mathbb{E}\left[ {\sin {\theta _1}  \sin {\theta _2}} \right].
\end{aligned}
\end{equation}
Therefore, the correlation of the real part is equal to the correlation of the imaginary part in the angular domain, if the mild conditions in (13) and (15) hold.
\end{appendices}

\ifCLASSOPTIONcaptionsoff
  \newpage
\fi

%

%






\end{document}